\documentclass{aa}  
\usepackage[normalem]{ulem}
\usepackage{xcolor}
\usepackage{graphicx}
\usepackage{txfonts}
\usepackage{lipsum}
\usepackage{subcaption}
\usepackage{lscape}
\usepackage{placeins}
\makeatletter
\AtBeginDocument{\nolinenumbers}

\begin{document}
%\nolinenumbers
   \title{Gaussian process analysis of type-B quasiperiodic oscillations in the black hole X-ray binary MAXI~J1348--630}
   \titlerunning{GP Modeling of Type-B QPOs in MAXI~J1348--630}

   \author{Yiran Wang\inst{1}
        \and Ruican Ma\inst{2}
        \and Haiyun Zhang\inst{3}
        \and Dahai Yan\inst{4}
            }
   \institute{School of Physics and Technology, Wuhan University, Wuhan, Hubei, 430072, People's Republic of China\\
            \and School of Physics and Astronomy, University of Southampton, Highfield, Southampton, SO17 1BJ, UK\\
            \email{R.Ma@soton.ac.uk}
            \and Yunnan Key Laboratory of Statistical Modeling and Data Analysis, Yunnan University, Kunming, Yunnan, 650091, People's Republic of China\\
            \and Department of Astronomy, Key Laboratory of Astroparticle Physics of Yunnan Province, Yunnan University, Kunming, Yunnan, 650091, People's Republic of China\\
            \email{yandahai@ynu.edu.cn}
        }

   \date{Received May 17, 2025}

%nolinenumbers
\g@addto@macro\maketitle{\nolinenumbers}
\makeatother
 
  \abstract{
  We analyzed Insight-HXMT data of the black hole X-ray binary MAXI~J1348--630 during the type-B quasiperiodic oscillation (QPO) phase of its 2019 outburst. Using the Gaussian process method, we applied an additive composite kernel model consisting of a stochastically driven damped simple harmonic oscillator (SHO), a damped random walk (DRW), and an additional white noise (AWN) to data from three energy bands: low energy (LE; 1--10\,keV) band, medium energy (ME; 10--30\,keV) band, and high energy (HE; 30--150\,keV) band. 
  We find that for the DRW component, correlations on the timescale of $\tau_{\rm DRW}\sim10\,$s are absent in the LE band, while they persist in the ME and HE bands over the full duration of the light curves. This energy-dependent behavior may reflect thermal instabilities, with the shorter correlation timescale in the disk compared to the corona. 
  Alternatively, it may reflect variable Comptonizations of seed photons from different disk regions. Inner-disk photons are scattered by a small inner corona, producing soft X-rays. Outer-disk photons interact with an extended, jet-like corona, resulting in harder emission.
  The QPO is captured by an SHO component with a stable period of $\sim 0.2$\,s and a high quality factor of $\sim 10$. 
  The absence of significant evolution with energy or time of the SHO component suggests a connection between the accretion disk and the corona, which may be built by coherent oscillations of disk-corona driven by magnetorotational instability.
  The AWN components are present in all the three-band data and dominate over the DRW and SHO components.
  We interpret the AWN as another fast DRW with its $\tau_{\rm DRW} < 0.01$\,s. 
  It may trace high-frequency fluctuations that occur in both the inner region of the accretion disk and the corona.
  Overall, our work reveals a timescale hierarchy in the coupled disk-corona scenario: fast DRW < SHO < disk DRW < corona DRW.
  }

   \keywords{Methods: data analysis --
                Methods: statistical --
                X-rays: binaries
               }

   \maketitle

%%%%%%%%%%%%%%%%%%%%%%%%%%%%%%%%%%%%%%%%%%%%%%%%%%%%%%%%%%%%%%%%%%%%%%%%%%
\section{Introduction}
Black hole X-ray binaries (BHXBs) are compact binary systems, each composed of a stellar-mass black hole and a companion star. Most low-mass BHXBs are transient sources, spending the majority of their time in a low-luminosity quiescent state and occasionally undergoing dramatic outbursts that last from weeks to months  \citep[e.g.,][]{tanaka1996x}. These outbursts are generally thought to be triggered by accretion disk instabilities \citep{lasotaDiscInstabilityModel2001}. A typical BHXB outburst can be characterized by distinct spectral and timing properties, allowing classification into several accretion states \citep[see ][for a review]{mendezEXOSATDataGX1997a, mcclintock2006compact}. The outburst usually begins in the low-hard state (LHS), where the emission is dominated by a hard Comptonization component, and the accretion disk is believed to be truncated at tens to hundreds of gravitational radii \citep[$R_{\rm g}$;][]{doneModellingBehaviourAccretion2007}. However, some studies reporting the detection of a broad iron line suggest that the accretion disk may not be truncated in the LHS \citep[e.g.,][]{2010MNRAS.402..836R, 2006ApJ...653..525M}. As the outburst evolves, the contribution from the soft thermal disk component increases, and the disk extends inward to reach the innermost stable circular orbit \citep[ISCO;][]{Esin_1997}, marking the transition to the high-soft state (HSS). The intermediate state (IMS) serves as a transitional phase between the LHS and the HSS and can be further subdivided into the hard intermediate state (HIMS) and the soft intermediate state (SIMS), with the SIMS exhibiting a softer spectrum than the HIMS. Throughout a typical outburst, the evolution of the source can be traced in a model-independent way using the hardness-intensity diagram (HID), in which BHXBs typically follow a characteristic "q"-shaped track \citep[e.g.,][]{Homan_2001, homanEvolutionBlackHole2005}.\par

Fast X-ray variability is another prominent feature observed during outbursts of BHXBs \citep[see][for a review]{1988SSRv...46..273L, ingramReviewQuasiperiodicOscillations2019b}. The most significant and extensively studied timing features are quasiperiodic oscillations (QPOs), which typically appear as one or more narrow peaks in the power spectral density (PSD) \citep[e.g.,][]{motchSimultaneousXrayOptical1983, vanderklisQuasiperiodicOscillationsNoise1989}. Low-frequency QPOs (LFQPOs) are commonly classified into three types, A, B, and C, based on the shape and strength of the noise component in the PSD and the fractional root mean square (rms) amplitude \citep[e.g.,][]{Wijnands1999, remillardCharacterizingQuasiperiodicOscillation2002, casellaABCLowfrequencyQuasiperiodic2005a, ingramReviewQuasiperiodicOscillations2019b}.
In particular, type-B QPOs are predominantly observed during the SIMS. They typically have centroid frequencies around $5-6$\,Hz, quality factors ($Q=\nu/\rm FWHM$, where $\rm FWHM$ is full width at half maximum) up to 6, and relatively high fractional rms amplitudes of $\sim 2-4$\% \citep{casellaABCLowfrequencyQuasiperiodic2005a}. Type-B QPOs are often suggested to be linked to jet activity \citep[e.g.,][]{Giannios2004, soleriTransientLowfrequencyQuasiperiodic2007, Kylafis2008, stevensPhaseresolvedSpectroscopyType2016,maVariableCoronaTransition2023a}. 
For instance, \citet{2020A&A...640L..16K} proposed a quantitative jet model to explain the origin of type-B QPOs, while \citet{liuTransitionsOriginTypeB2022} suggested that these QPOs may be related to the precession of a weak jet within a tilted disk-jet configuration located relatively close to the black hole. 
The noise component observed during the SIMS is typically weak and dominated by red noise. This red noise generally has a fractional rms amplitude of only a few percent and is confined to low frequencies \citep[typically below 0.1\,Hz;][]{casellaABCLowfrequencyQuasiperiodic2005a}. Despite its frequent occurrence, the physical origin of this noise component remains poorly understood.\par

MAXI~J1348--630 is a BHXB, first observed in the X-ray band with the Monitor of All-sky X-ray Image \citep[MAXI;][]{matsuokaMAXIMissionISS2009} on January 26, 2019 \citep{yatabeMAXIGSCDiscovery2019}. Based on the spectral and timing properties observed with the Neutron Star Interior Composition Explorer \cite[NICER;][]{gendreauNeutronStarInterior2016}, \citet{sannaNICERIdentificationMAXI2019} suggested that the compact object in this system is a black hole. MAXI~J1348--630 exhibited a typical outburst and was followed by four subsequent hard-state re-brightenings \citep{alyazeediRebrighteningMAXIJ13486302019,yatabeMAXIGSCDiscovery2019,russellOpticalBrighteningMAXI2019,carotenutoMeerKATSwiftXRT2020,pirbhoyXBNEWSDetectionNew2020,shimomukaiMAXIJ1348630MAXI2020,zhangRebrighteningDecayingMAXI2020}. \citet{lamerGiantXrayDust2021} analyzed the source's X-ray dust scattering halo, estimating its distance to be $3.39\pm{0.34}$\,kpc, and their X-ray spectral modeling further indicates a black hole of $11\pm{2}\,M_{\odot}$.
The inclination angle of the jet was constrained to ${29.3^{+2.7}_{-3.2}}^{\circ}$ by modeling its motion with a dynamical external shock framework \citep{carotenutoModellingKinematicsDecelerating2022}. Additionally, employing a joint continuum and reflection fitting technique with Insight-HXMT observations, \citet{guan2024estimating} estimated the black hole spin to be $0.79\pm{0.13}$.\par

Since the discovery of MAXI~J1348--630 in 2019, the timing properties of MAXI~J1348--630 have been extensively studied. For instance, \citet{jitheshBroadbandSpectralTiming2021} used AstroSat and NICER observations to detect a type-C QPO at approximately 0.9 Hz and a type-A QPO at around 6.9\,Hz. 
\citet{2022MNRAS.514.2839A} investigated the properties of type-C QPOs during both the main outburst and the re-flare state, and explained their rms spectra and lag spectra within the framework of Comptonization radiation. More recently, \citet{alabartaGeometryComptonizationRegion2025a} studied the Comptonization region in MAXI~J1348--630 during the LHS and HIMS of both the main outburst and re-flare. By examining the fractional rms and lag spectra of type-C QPOs, they concluded that a sudden increase in the phase-lag frequency spectrum and a sharp drop in the coherence function during the decay phase are likely caused by the type-C QPOs. 
As for the type-B QPOs in MAXI~J1348--630, \citet{zhangNICERUncoversTransient2021} conducted a spectral-timing analysis of four NICER observations, focusing on the rapid appearance and disappearance of type-B QPOs in the 0.5--10\,keV energy band. Subsequently, \citet{liuTransitionsOriginTypeB2022} analyzed Insight-HXMT data to investigate the origin of type-B QPOs, suggesting a possible link with jet precession. 
\citet{garciaTwocomponentComptonizationModel2021} and \citet{2022MNRAS.515.2099B} successfully explained the rms and lag spectra properties of type-B QPO, particularly at high energies, using a model with two physically connected Comptonization regions. \par

In this paper, we present a detailed analysis of the timing properties of MAXI~J1348--630 during its type-B QPO phase.
The Gaussian process (GP) method was used to identify patterns in the X-ray data and to further explore underlying physical mechanisms. 
The structure of the paper is as follows: Section 2 presents the data reduction. Section 3 provides a brief overview of the GP framework. Section 4 presents the results of the model fitting. Section 5 discusses the implications of our findings.

%%%%%%%%%%%%%%%%%%%%%%%%%%%%%%%%%%%%%%%%%%%%%%%%%%%%%%%%%%%%%%%%%%%%%%%%%%

\section{Observations and data reduction}
Insight-HXMT is China's first X-ray astronomical satellite, which aims at observing X-ray sources within a broad energy band of 1--250\,keV. The satellite is equipped with three instruments: the High Energy X-ray telescope (HE; 20--250\,keV) \citep{liuHighEnergyXray2020}, the Medium Energy X-ray telescope (ME; 5--30\,keV) \citep{caoMediumEnergyXray2020}, and the Low Energy X-ray telescope (LE; 1--15\,keV) \citep{chenLowEnergyXray2020,zhangOverviewHardXray2020a}. In this work, we focus on the SIMS observations of MAXI~J1348--630 obtained with Insight-HXMT. The ObsIDs used are P0214002011, P0214002012, and P0214002015--P0214002017, spanning MJD 58522.6--58527.9. \par

Using the Insight-HXMT Data Analysis software (HXMTDAS, v2.06), we performed the data reduction by applying the following selection criteria: an elevation angle greater than $10^{\circ}$, a geomagnetic cut-off rigidity greater than 8\,GV, a pointing offset angle less than or equal to $0.04^{\circ}$, and time intervals at least 300\,seconds before and after the South Atlantic Anomaly (SAA) passage.
The selected energy bands are 1--10\,keV for LE, 10--30\,keV for ME, and 30--150\,keV for HE.

%%%%%%%%%%%%%%%%%%%%%%%%%%%%%%%%%%%%%%%%%%%%%%%%%%%%%%%%%%%%%%%%%%%%%%%%%%
\section{Data analysis}
\subsection{Gaussian process method}
The GP consists of a class of probabilistic models with strong predictive capabilities for continuous stochastic processes, which has been used in time-domain astronomy \citep{aigrainGaussianProcessRegression2023}. 
As an extension of the Gaussian distribution to the multivariate case, GP consists of a mean function $\mu_{\mathbf{\theta}}(\mathbf{x})$ and a covariance (or kernel) function $k_{\mathbf{\alpha}}(\mathbf{x}_n, \mathbf{x}_m)$ parameterized by $\mathbf{\theta}$ and $\mathbf{\alpha}$, where $\mathbf{x}_n$ and $\mathbf{x}_m$ represent the \( n \)-th and \( m \)-th data points, respectively. The mean function describes the expected value of the multivariate Gaussian distribution. The covariance function models the variance along each dimension and determines how the different random variables are correlated. In practice, the mean function, $\mu$, is defined as the mean value of the time series, and the key point in building the GP model lies in the choice of the kernel function. Applying the GP model to the data, Markov chain Monte Carlo (MCMC) methods are often used to sample the parameter space and quantify the parameters' uncertainties.\par

The GP method has been used to study the QPOs and the stochastic variability of active galactic nuclei \citep{zhangEvidenceMagnetogravitationalProcesses2025,zhangGaussianProcessModeling2023,zhangQuasiperiodicOscillationGray2021}.
Here, we list the key formulas in the GP package of \texttt{celerite}, presented by \citet{foreman-mackeyFastScalableGaussian2017}.
For the time series, 
the \texttt{celerite} covariance function
is given by \citep[see also][]{PhysRevLett.74.1060}

\begin{equation}
    k_{\alpha}(t_{nm}) = (\sigma^e_n)^2 \delta_{nm} + \sum_{j=1}^{J} a_j \exp(-c_j t_{nm}).\label{var}
\end{equation}
In this notation, \( n \) and \( m \) indicate the \( n \)-th and \( m \)-th time points within the total $N$ time points, with \( t_{nm} \) denoting the time interval between these two points.
Here $\{(\sigma^e_n)^2\}_{n=1}^N$ are the reported measurement uncertainties, $\delta_{nm}$ is the Kronecker delta, and $\boldsymbol{\alpha} = (a, c)$ is the covariance function parameter vector.

When introducing complex parameters into the second term of Eq.~(\ref{var}) and rewriting the exponentials, a more general covariance function is obtained:
\begin{equation}
    k_{\mathbf{\alpha}}(t_{nm}) = (\sigma^e_n)^2 \delta_{nm} + \sum_{j=1}^{J}
     \left[ a_j e^{-c_j t_{nm}} \cos(d_j t_{nm}) 
    + b_j e^{-c_j t_{nm}} \sin(d_j t_{nm}) \right].\label{newvar}
\end{equation}
Each term in the sum is a damped oscillatory component, referred to as a "\texttt{celerite} term." The parameter $c_j$ controls the exponential decay (damping) rate of the oscillation with the angular frequency $d_j$. The parameters $a_j$ and $b_j$ are amplitude parameters that set the relative weighting of the cosine and sine components, respectively.

Performing the Fourier transform to this covariance function, 
the PSD is then obtained:
\begin{equation}
    S(\omega) = \sum_{j=1}^{J} \sqrt{\frac{2}{\pi}} \frac{(a_j c_j + b_j d_j)(c_j^2 + d_j^2) + (a_j c_j - b_j d_j)\omega^2}{\omega^4 + 2(c_j^2 - d_j^2)\omega^2 + (c_j^2 + d_j^2)^2}.
    \label{psd1}
\end{equation}

We can relate the dynamics of a stochastically driven damped simple harmonic oscillator (SHO) to the GP. The standard form for a damped SHO driven by a random (white noise) forcing $\epsilon(t)$ is 
\begin{equation}
    \left[ \frac{d^2}{dt^2} + \frac{\omega_0}{Q} \frac{d}{dt} + \omega_0^2 \right] y(t) = \epsilon(t),\label{SHO_t}
\end{equation}
where $\omega_0$ is the natural frequency of the undamped oscillator and $Q$ is its quality factor of the oscillator.
The PSD of this process is
\begin{equation}
    S(\omega) = \sqrt{\frac{2}{\pi}} \frac{S_0 \omega_0^4}{(\omega^2 - \omega_0^2)^2 + \frac{\omega_0^2 \omega^2}{Q^2}}, \label{SHO}
\end{equation}
where $S(\omega_0)=\sqrt{2/\pi} S_0 Q^2$.

To match the SHO PSD with the PSD in Eq.~(\ref{psd1}),
we set $a_j=S_0\omega_0Q$, $b_j=a_j/\sqrt{4Q^2-1}$,
$c_j=\omega_0/(2Q)$, and $d_j=c_j\sqrt{4Q^2-1}$.
Putting these parameters in Eq.~(\ref{newvar}), 
the corresponding SHO kernel was obtained.
The quality factor $Q$ controls the oscillation modes of the SHO kernel. For a high-quality oscillator with $Q\gg1$, 
the SHO kernel was reduced to 
\begin{equation}
    k_{\text{SHO}}(t_{nm}) \approx S_0 \omega_0 Q e^{-\frac{\omega_0 t_{nm}}{2 Q}} \text{cos}(\omega_0 t_{nm}).\label{SHO_kernel}
\end{equation}
When $Q\ll1$, the damping term controls the kernel, leading to the absence of oscillations, which is referred to as the over-damped mode. For $Q\sim0.5$, the system is in the critical damped state. 
The parameter $Q$ can also be expressed as $Q=\omega_0 / 4 \pi \Gamma$, where $\Gamma$ is the half width at half maximum of a peak in the PSD.

The variance of this term is the integral of Eq.~(\ref{SHO}) over the entire frequency band \citep{osullivanModellingStochasticQuasiperiodic2024}:
\begin{equation}
    {\rm VAR}_{\text{SHO}}= \int_{- \infty}^{\infty} \sqrt{\frac{2}{\pi}}\, 
\frac{S_{0} \,\omega_{0}^{4}}{\left(\omega^{2} - \omega_{0}^{2}\right)^{2}
+ \frac{\omega_{0}^{2}\,\omega^{2}}{Q^{2}}}
\, d\omega = S_0 \omega_0 Q. \label{varsho}
\end{equation}
Also, the three parameters -- $\omega_0$, $S_0$, and $Q$ -- can be reformulated and defined as three new parameters, including the period $\rho$, damping timescale $\tau$, and standard deviation $\sigma$:
\begin{equation}
    \rho \equiv \frac{2\pi}{\omega_0}, \quad \tau \equiv \frac{2Q}{\omega_0}, \quad  \sigma \equiv \sqrt{S_0 \omega_0 Q}. \label{SHO_newpara}
\end{equation}

Another particular kernel is called the damped random walk (DRW).
The DRW kernel is related to the Ornstein-Uhlenbeck process \citep{PhysRev.36.823} in physics \citep[see also][]{1992ApJ...398..169R, PhysRevLett.74.1060}. 
Setting $b_j$ and $d_j$ to zero in Eq.~(\ref{newvar}), we obtain the DRW covariance function,
\begin{equation}
    k_j(t_{nm}) = a_j e^{-c_j t_{nm}}.\label{DRW1}
\end{equation}
Equivalently, Eq.~(\ref{DRW1}) can be written as\begin{equation}
    k_{\text{DRW}}(t_{nm})=\sigma_{\rm DRW}^2 e^{-\frac{t_{nm}}{\tau_{\rm DRW}}}, \label{DRW2}
\end{equation}
where $\sigma_{\rm DRW}=\sqrt{a_j}$ and $\tau_{\rm DRW}=c_j^{-1}$. 
The PSD of this process is
\begin{equation}
    S(\omega)= \frac{2\sigma_{\rm DRW}^2 \tau_{\rm DRW}}{1+\tau_{\rm DRW}^2 \omega^2}.\label{DRW_PSD}
\end{equation}
The DRW model is parameterized by the damping timescale $\tau_{\rm DRW}$ and the standard deviation $\sigma_{\rm DRW}$. The parameter $\tau_{\rm DRW}$ describes the rate at which the correlation of the data decays over time. The parameter $\sigma_{\rm DRW}$ is proportional to the asymptotic amplitude of the structure function when $t_{nm} \gg \tau_{\rm DRW}$, where the structure function describes the rms magnitude difference as a function of the time lag between observations \citep{zuQuasarOpticalVariability2013,macleodModelingTimeVariability2010}.
The damping timescale $\tau_{\rm DRW}$ corresponds to the break frequency $f_b=1/(2\pi\,\tau_{\rm DRW})$, where the PSD index changes from $2$ to $0$ from high frequencies to low frequencies.

The corresponding variance is given by \citep{osullivanModellingStochasticQuasiperiodic2024}
\begin{equation}
   {\rm VAR}_{\text{DRW}} = \int_{- \infty}^{\infty} 
\frac{2\sigma_{\rm DRW}^2 \tau_{\rm DRW}}{1 + \tau_{\rm DRW}^2 \omega^2} \, d\omega = \sigma_{\rm DRW}^2 \label{vardrw}.
\end{equation}
\par

The white noise associated with the Poisson noise is accounted for by the first term on the right-hand side of Eq.~(\ref{var}) \citep{foreman-mackeyFastScalableGaussian2017,osullivanModellingStochasticQuasiperiodic2024, zhangEvidenceMagnetogravitationalProcesses2025}. However, an additional white noise (AWN) term is sometimes required to improve the fitting \citep[e.g.,][]{2021Sci...373..789B,Zhang_2025}. This term is expressed as
\begin{equation}
    k_{\text{AWN}}(t_{nm})=\sigma_n^2 \delta_{nm},
\end{equation}
such that its variance is $\sigma_n^2$ and the vertical value of this white noise component in the PSD is $\sigma_n^2 \times t_{\rm binsize}$, with $t_{\rm binsize}$ being the time resolution of the data points.
Note that, unlike the reported measurement uncertainties that vary from point to point, the AWN term contributes equally to all data points.
Because the two contributions are uncorrelated, they simply add on the diagonal of the covariance matrix.

\subsection{Fitting procedure}
To fit the light curves from MAXI~J1348--630 during the outburst and obtain the posterior distribution of parameters, we used the powerful GP Python package \texttt{celerite} \citep{foreman-mackeyFastScalableGaussian2017}, 
and the MCMC fitting technique.  
The GP kernels remain valid under addition, multiplication, and convolution \citep{williams2006gaussian}. Here we used the \texttt{celerite} kernels and their additive combinations.

\texttt{celerite} has the ability to compute the characteristics of long-term variability \citep[e.g.,][]{2022ApJ...930..157Z} and short-term flare \citep[e.g.,][]{tangInsightsGaussianProcess2024,2025arXiv250220867Z}. In the program, log-uniform priors on each parameter were assumed. To ensure the stability of the results, we ran the MCMC sampler for 70,000 steps and dropped the initial 20,000 steps. The final 50,000 MCMC samples were used to construct the posterior distributions of the parameters and the PSD.\par

We also checked the standardized residuals, the autocorrelation function
(ACF), and the values of the Akaike information criterion corrected ($\rm AIC_c$) \citep[][]{sugiuraFurtherAnalysisData1978} to evaluate the quality of model fitting. If the distribution of the standardized residuals approximates a normal distribution and the ACF values of the standardized residuals fluctuate randomly within the 95\% confidence interval of the white noise, this indicates that the fitting is acceptable and the kernel is able to capture the underlying patterns in the data \citep[e.g.,][]{tangInsightsGaussianProcess2024}. For model comparison, the model with a lower $\rm AIC_c$ value was preferred.

%%%%%%%%%%%%%%%%%%%%%%%%%%%%%%%%%%%%%%%%%%%%%%%%%%%%%%%%%%%%%%%%%%%%%%%%%%

\section{Results}
\begin{figure}
    \centering
    \includegraphics[width=0.45\textwidth]{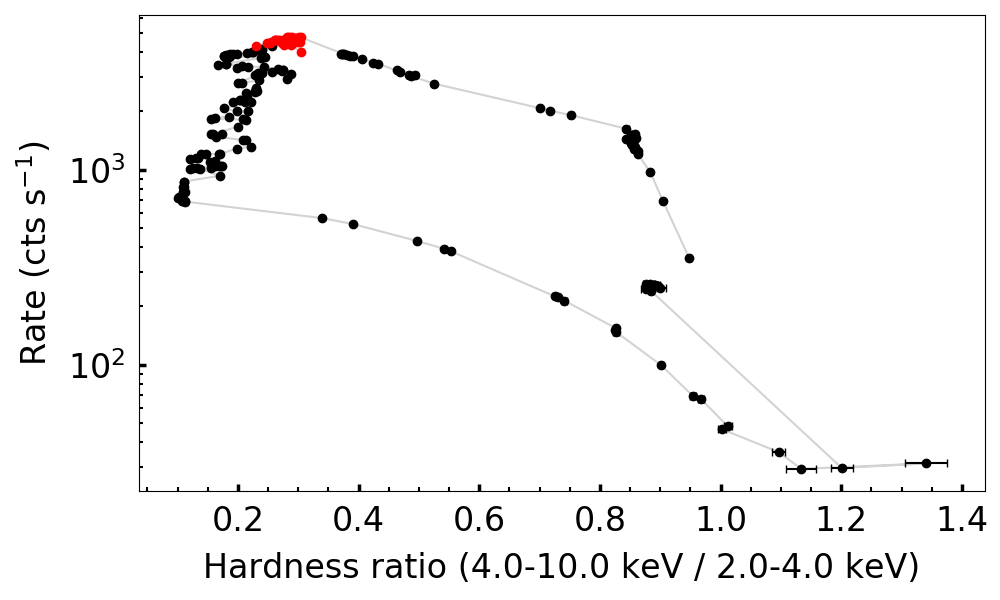}
    \captionsetup{singlelinecheck=off}
    \caption{HID of MAXI~J1348--630 during its 2019 outburst. The intensity is the count rate in the 2--10\,keV band, while the hardness ratio is the ratio of the count rate in the 4--10\,keV band to that in the 2--4\,keV band. Each point corresponds to one ExpID. The ExpIDs used in our work are marked in red.}\label{HID}
\end{figure}

% LE Qfree
\begin{figure*}
    \centering
    \begin{minipage}{0.48\textwidth}
        \centering
        \begin{subfigure}{\textwidth}
            \includegraphics[width=\textwidth]{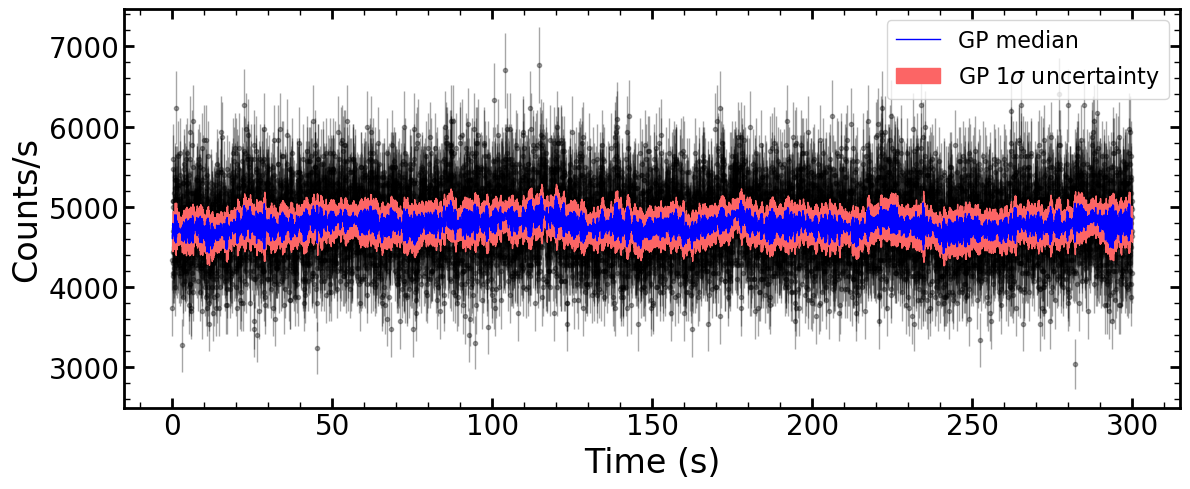}
        \end{subfigure}\\[0.55cm]
        \begin{subfigure}{\textwidth}
            \includegraphics[width=\textwidth]{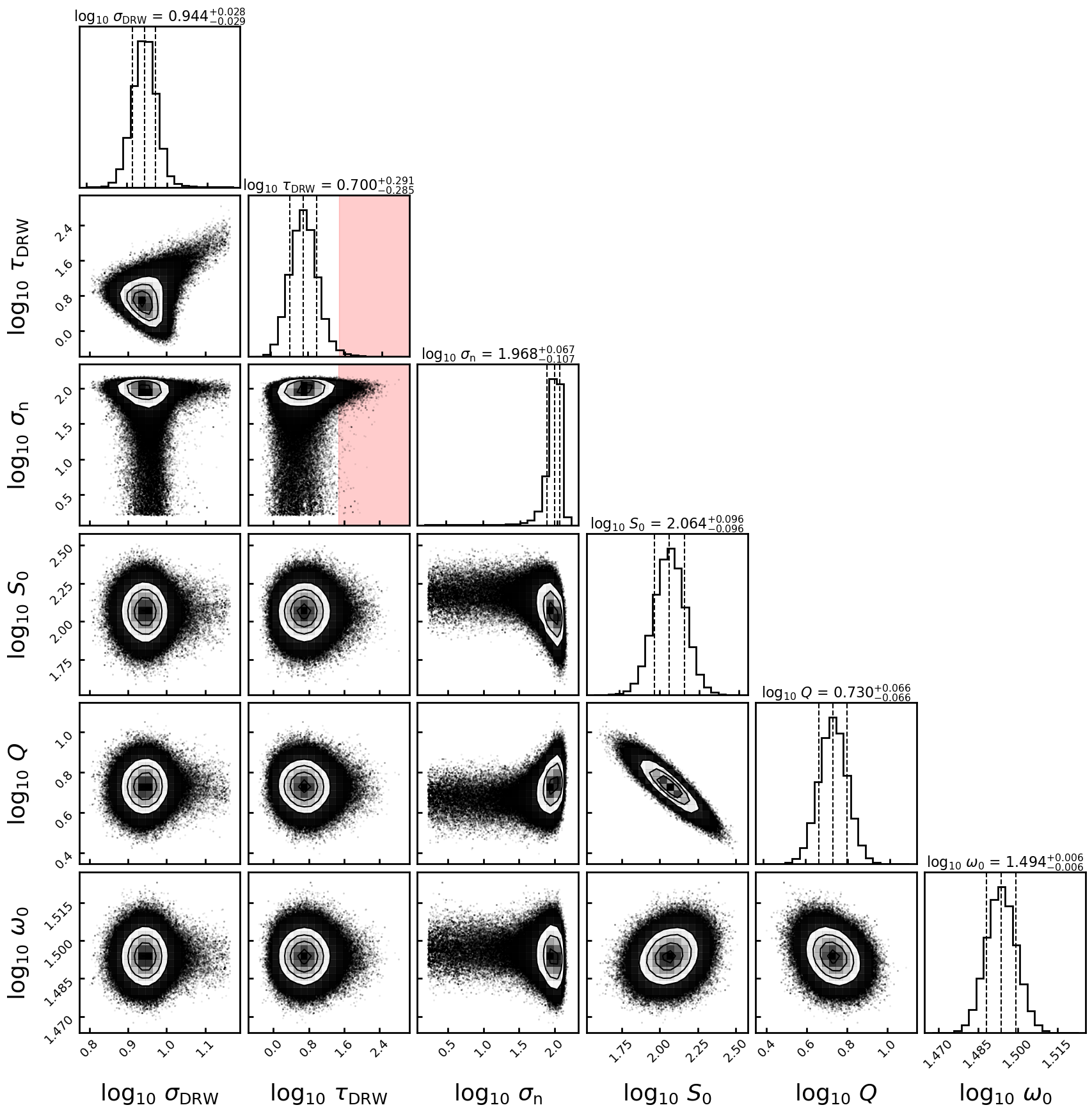}
        \end{subfigure}
    \end{minipage}
    \hspace{0.85cm}
    \begin{minipage}{0.45\textwidth}
        \centering
        \begin{subfigure}{\textwidth}
            \includegraphics[width=\textwidth]{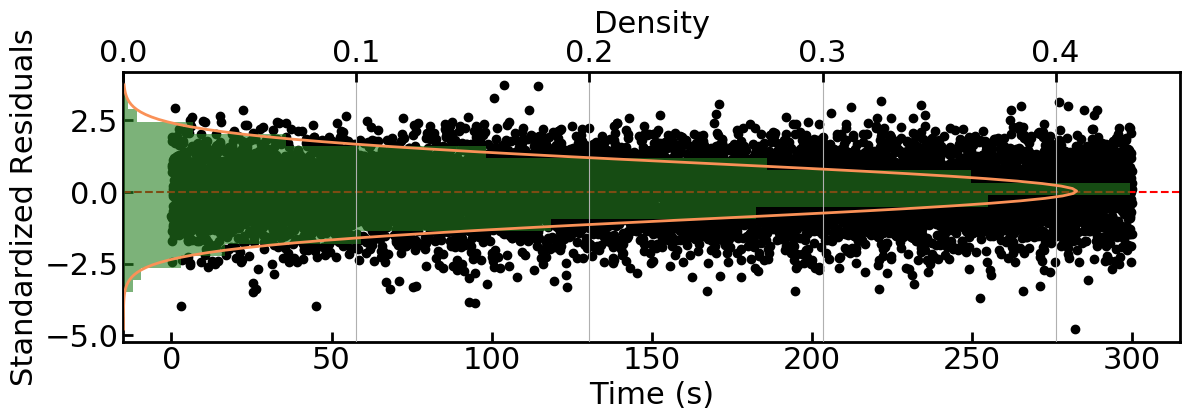}
        \end{subfigure}\\[1em]
        \begin{subfigure}{\textwidth}
            \includegraphics[width=\textwidth]{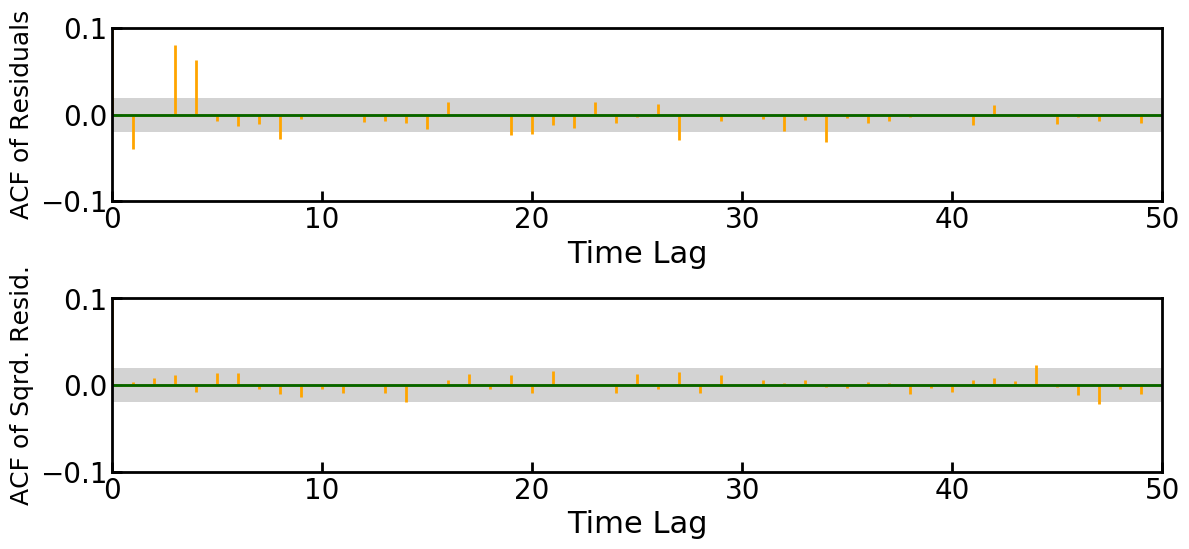}
        \end{subfigure}\\[1em]
        \begin{subfigure}{\textwidth}
            \includegraphics[width=1.\textwidth]{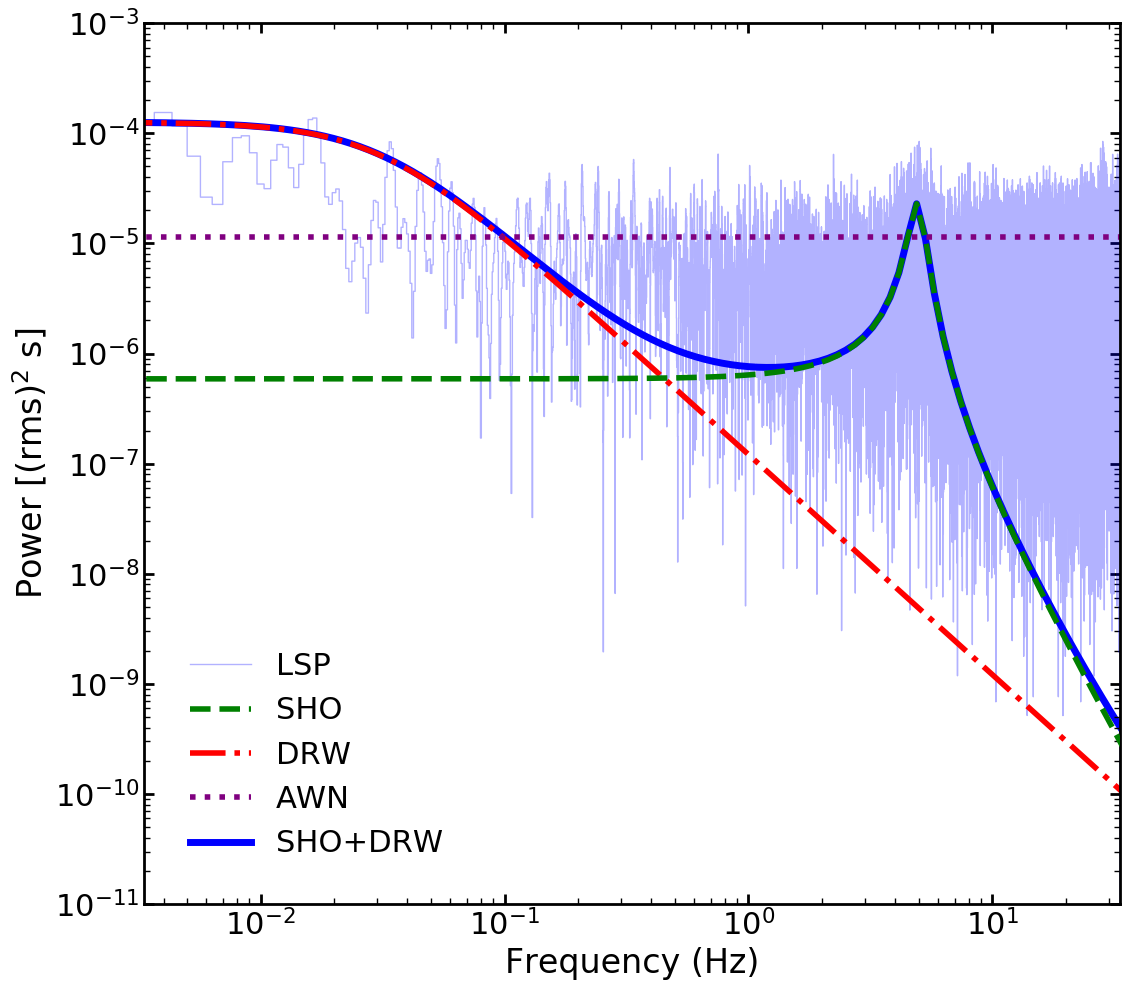}
        \end{subfigure}
    \end{minipage}
    \caption{LE fitting results when $Q$  is free. Left column, from top to bottom: Light curve fitting results (including the mean prediction from the GP model and the 1$\sigma$ uncertainty range corresponding to the total kernel) and parameter corner plot. Right column, from top to bottom: Residual distribution, ACF, and PSD. In PSD plot the blue line in the background represents the LSP, and the darker blue line is the total PSD of SHO (green line) and DRW (red line). The AWN term is plotted with the dotted purple line.}
    \label{LE_Qfree}
\end{figure*}

%  LE Q30
\begin{figure*}
    \centering
    \begin{minipage}{0.48\textwidth}
        \centering
        \begin{subfigure}{\textwidth}
            \includegraphics[width=\textwidth]{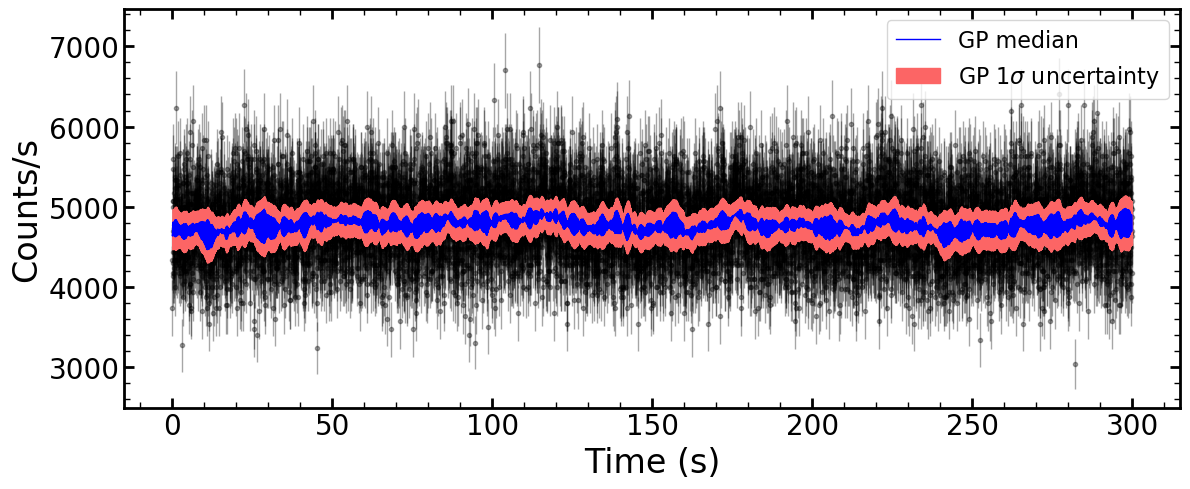}
        \end{subfigure}\\[0.55cm]
        \begin{subfigure}{\textwidth}
            \includegraphics[width=\textwidth]{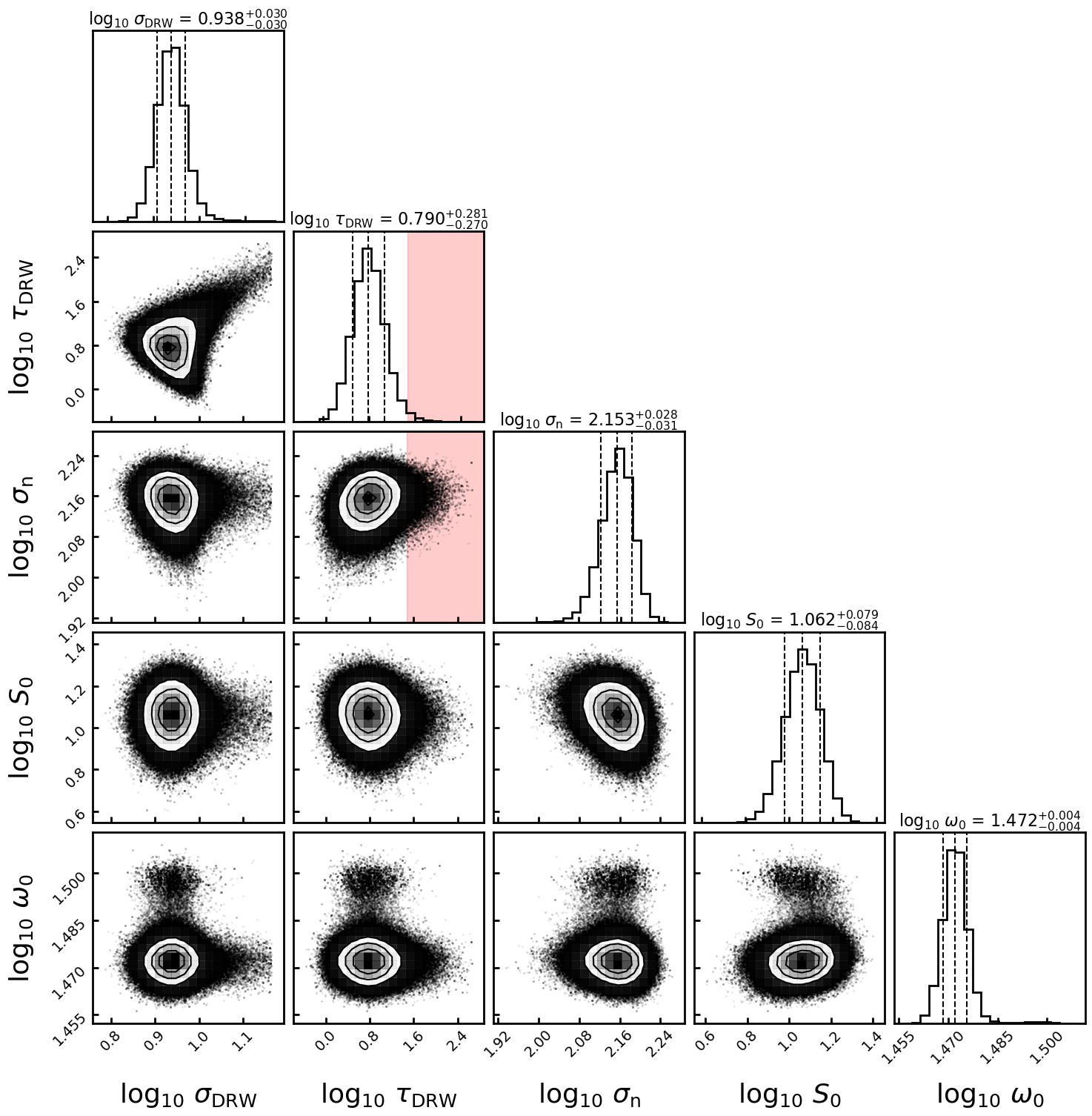}
        \end{subfigure}
    \end{minipage}
    \hspace{0.85cm}
    \begin{minipage}{0.45\textwidth}
        \centering
        \begin{subfigure}{\textwidth}
            \includegraphics[width=\textwidth]{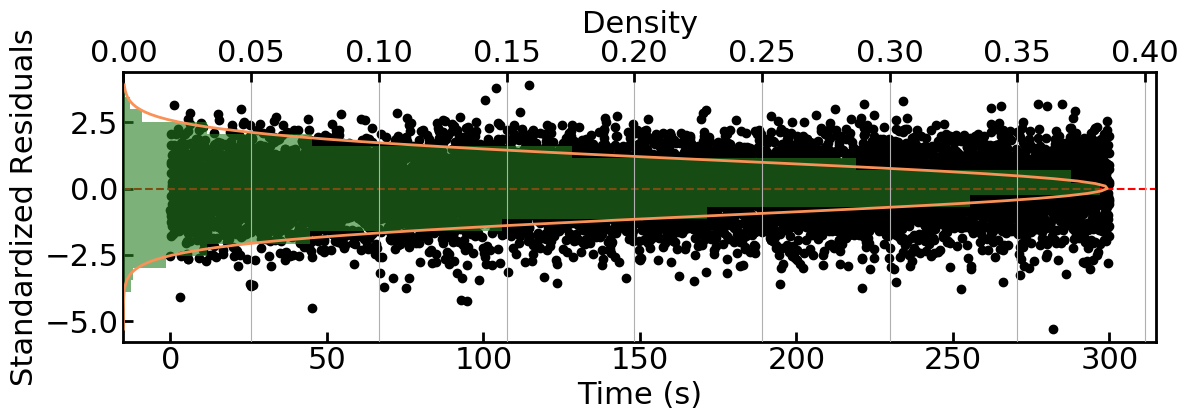}
        \end{subfigure}\\[1em]
        \begin{subfigure}{\textwidth}
            \includegraphics[width=\textwidth]{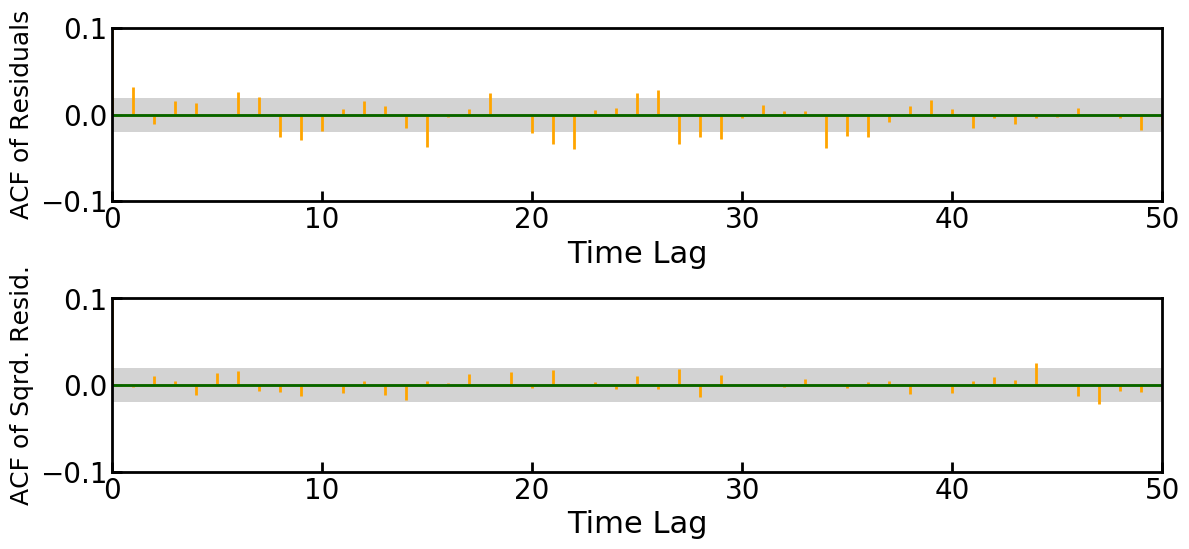}
        \end{subfigure}\\[1em]
        \begin{subfigure}{\textwidth}
            \includegraphics[width=1.\textwidth]{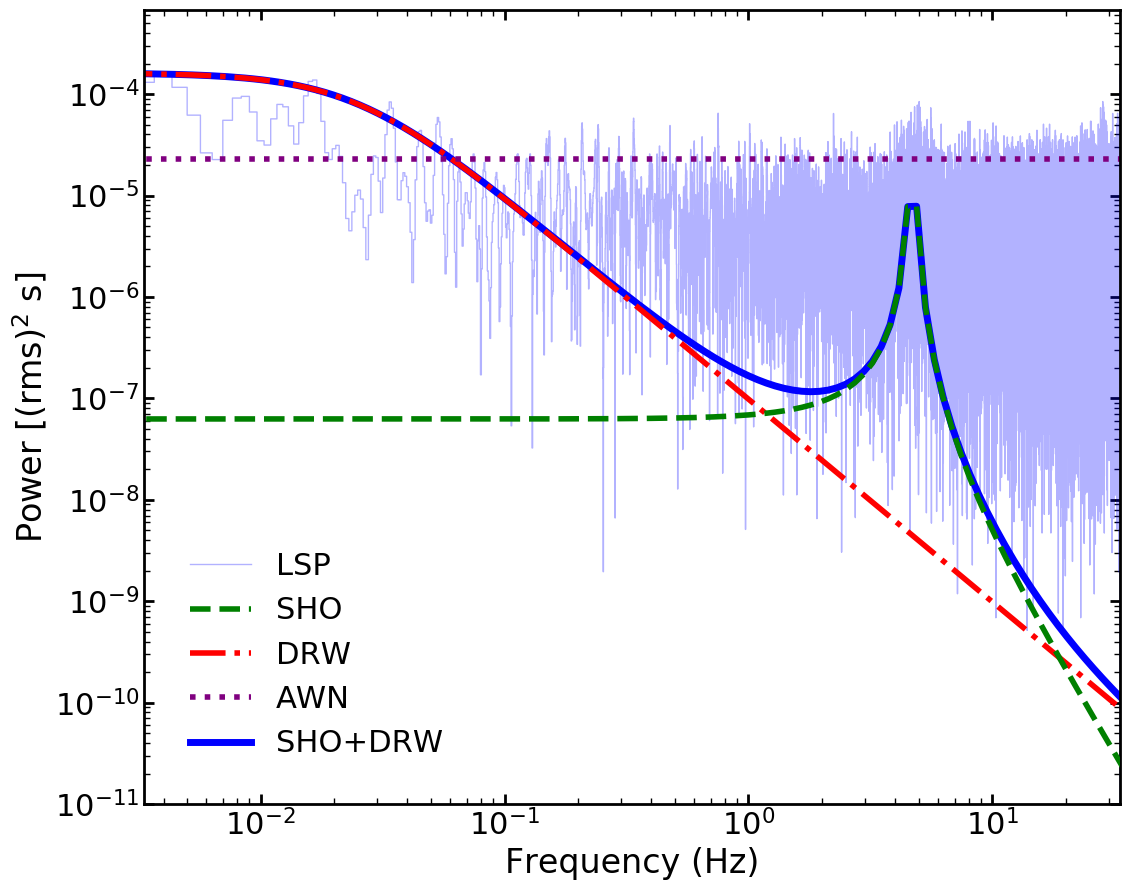}
        \end{subfigure}
    \end{minipage}
    \caption{Same as in Fig.~\ref{LE_Qfree} but $Q$ is fixed at 30.
    }
    \label{LE_Q30}
\end{figure*}

%HE
\begin{figure*}
    \centering
    \begin{minipage}{0.385\textwidth}
        \centering
        \begin{subfigure}{\textwidth}
            \includegraphics[width=\textwidth]{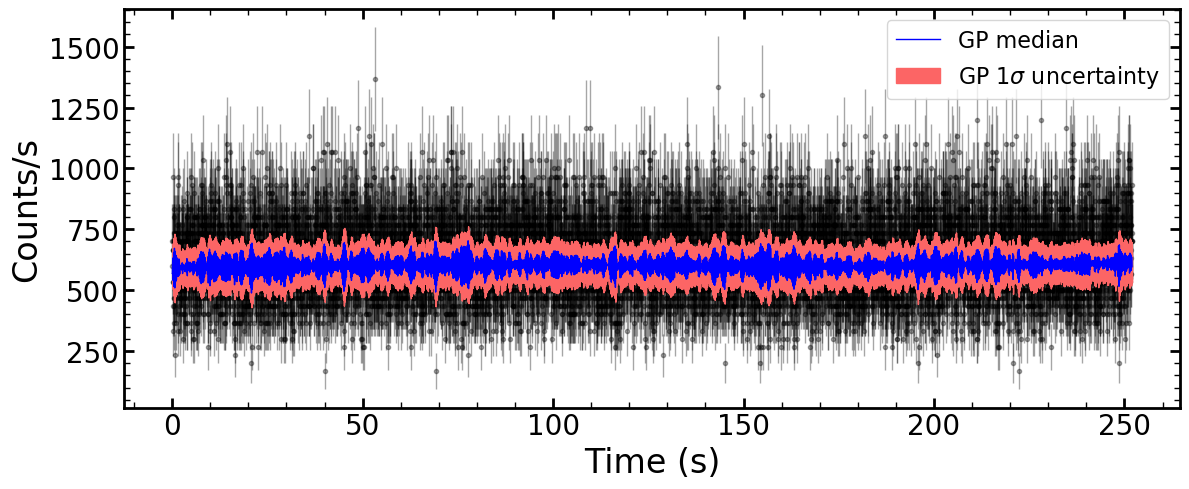}
        \end{subfigure}\\[0.1cm]
        \begin{subfigure}{\textwidth}
            \includegraphics[width=\textwidth]{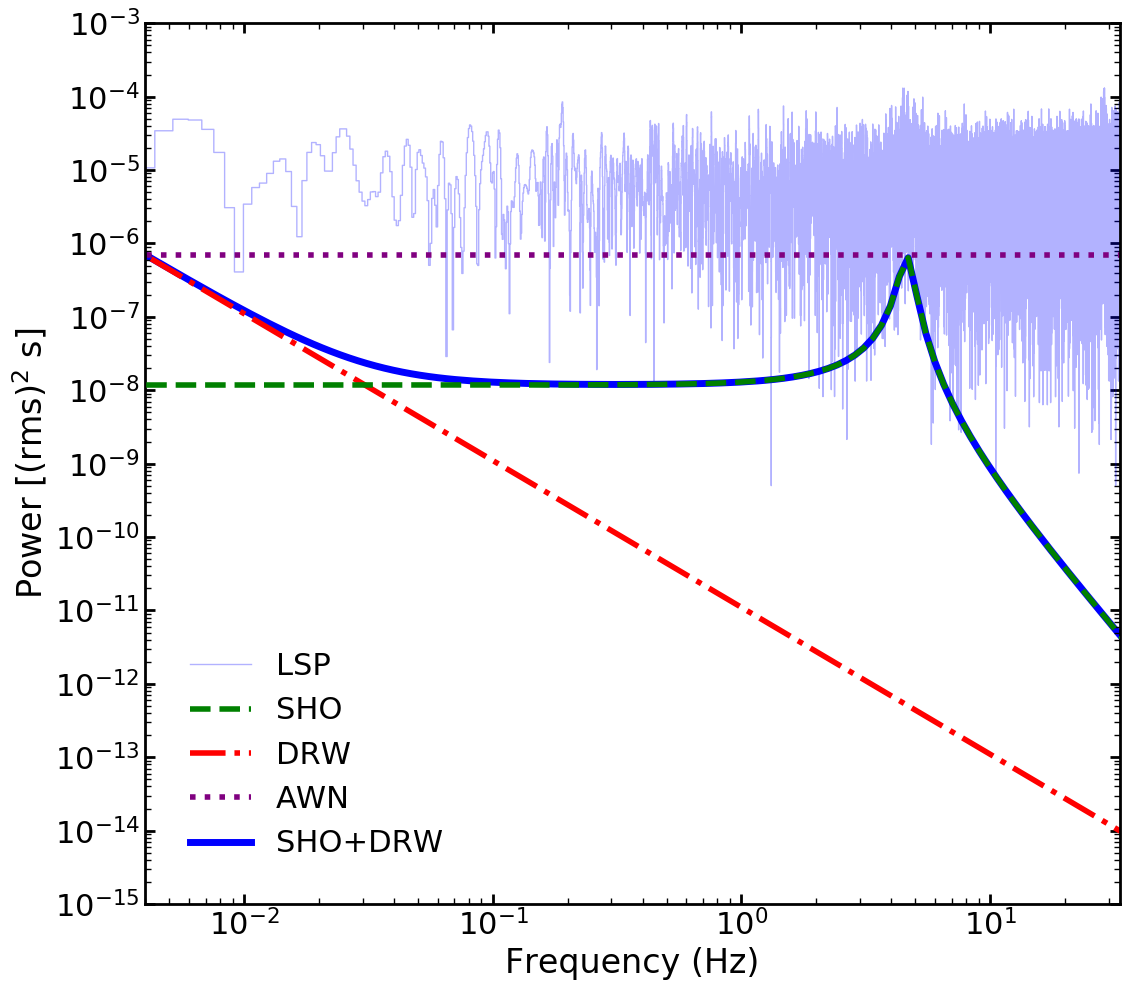}
        \end{subfigure}
    \end{minipage}
    \hspace{0.85cm}
    \begin{minipage}{0.5\textwidth}
        \centering
        \begin{subfigure}{\textwidth}
            \includegraphics[width=1.\textwidth]{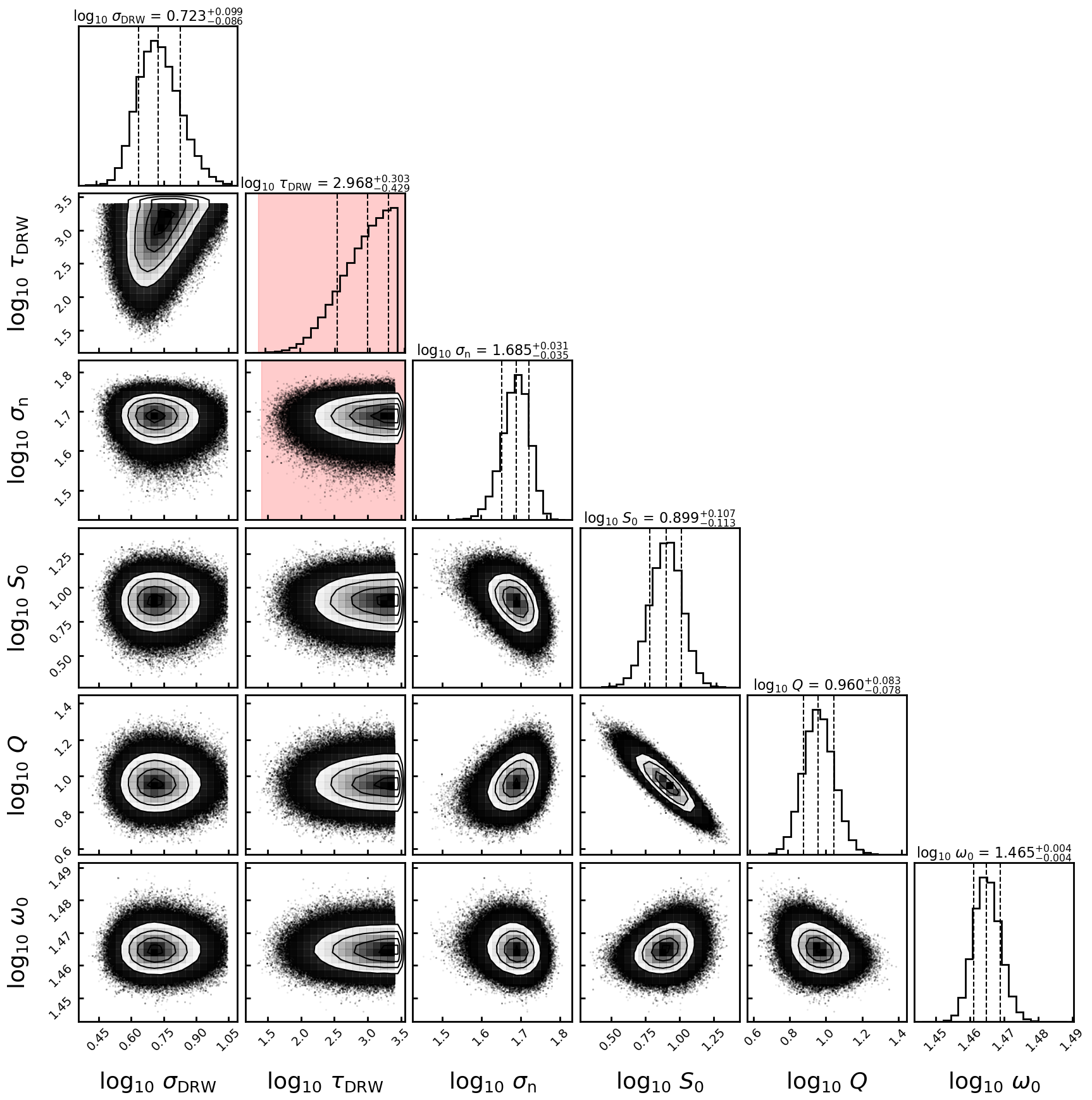}
        \end{subfigure}
    \end{minipage}
    \caption{HE fitting results when $Q$ is free. Left column, from top to bottom: Light curve fitting results (including the mean prediction from the GP model and the 1$\sigma$ uncertainty range corresponding to the total kernel) and PSD. Right column: Parameter corner plot. The ME fitting results are similar to the HE results, which are not shown. For both HE and ME, $\tau_{\rm DRW}$ is not effectively constrained.}
    \label{HE}
\end{figure*}

%compare PSD
\begin{figure*}
    \centering
    \begin{subfigure}[b]{0.4\textwidth}
        \includegraphics[width=\textwidth]{figures/good1101LE1/Q/psd_plot.png}
    \end{subfigure}
    \hspace{0.01\textwidth}
    \begin{subfigure}[b]{0.55\textwidth}
        \includegraphics[width=0.9\textwidth]{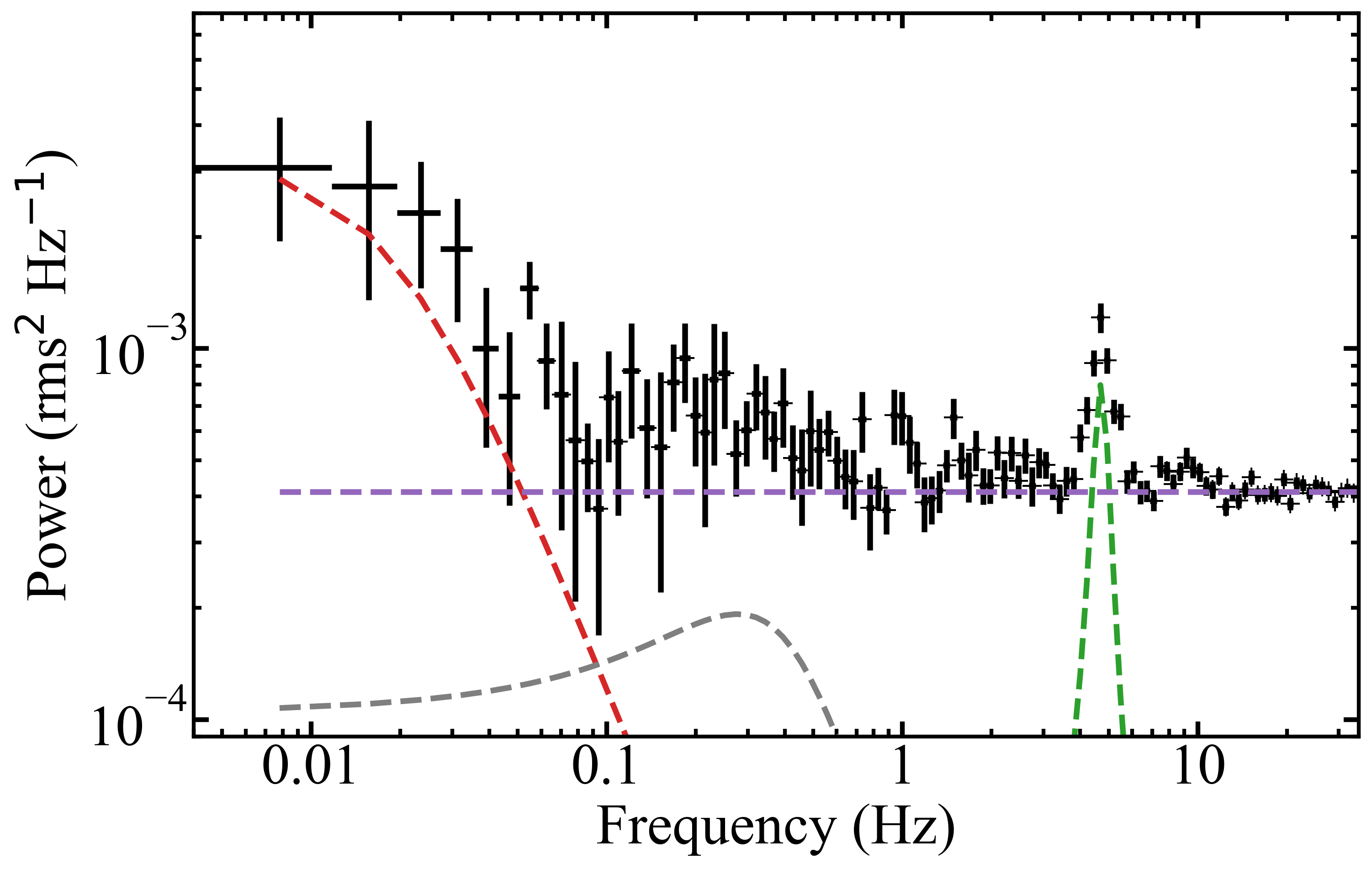}
    \end{subfigure}

    \caption{PSD obtained using the GP method (left panel) and the PSD computed with \texttt{powspec} (right panel) for the same ExpID in the LE energy band.}
    \label{good_pow}
\end{figure*}

%RMS
\begin{figure}
    \centering
    \includegraphics[width=0.47\textwidth]{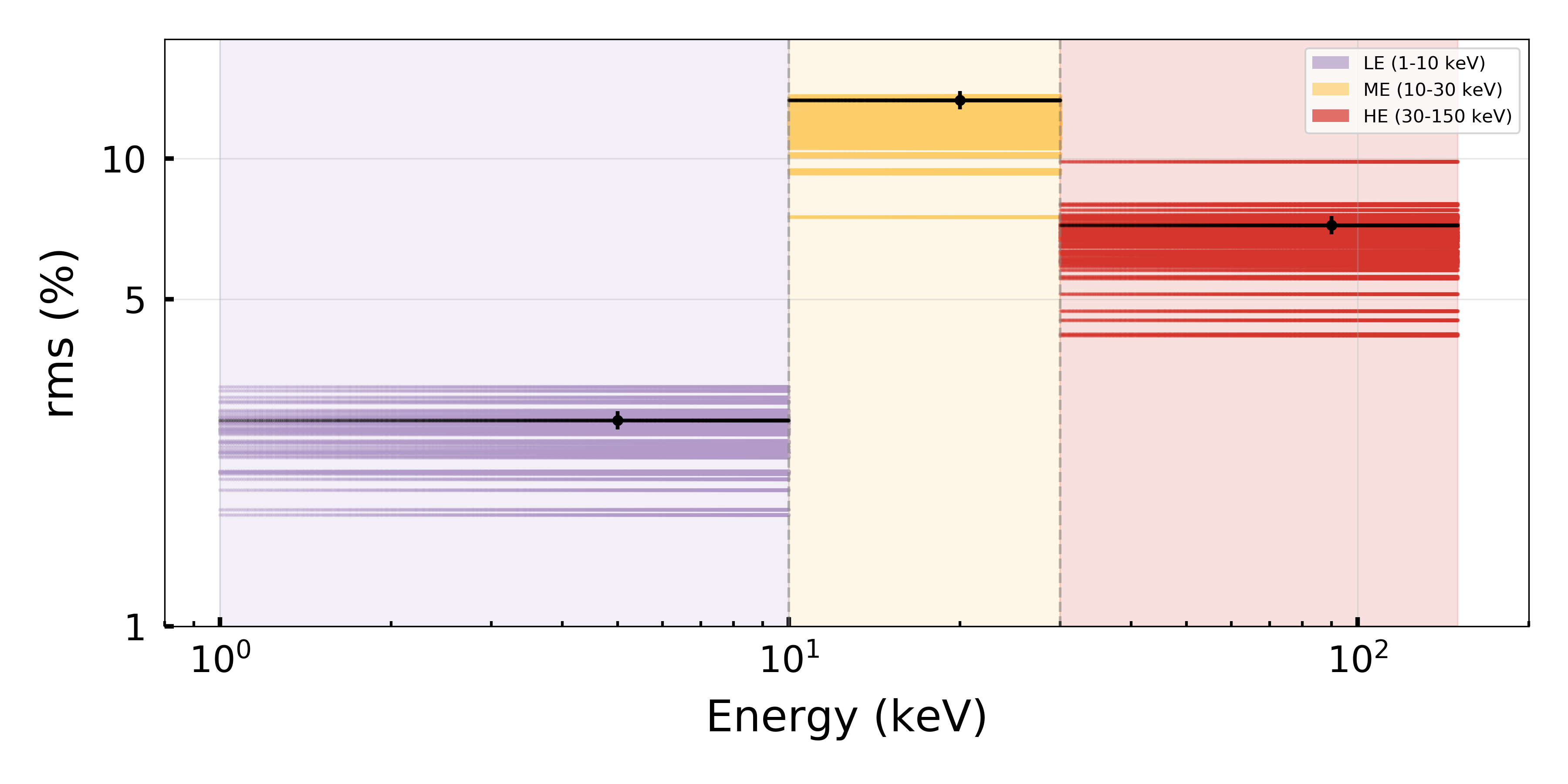}
    \caption{Fractional rms of the three energy bands (LE, ME, and HE), calculated by dividing the standard deviation of the SHO term by the mean count rate. The colors (purple, yellow, and red) correspond to the three discrete energy bands, with each line indicating the rms value for that GTI within the respective band. The three black lines correspond to the fractional rms of a representative ExpID calculated using the frequency-domain method.}\label{rms}
\end{figure}

%para
\begin{figure}
    \centering
    \includegraphics[width=0.47\textwidth]{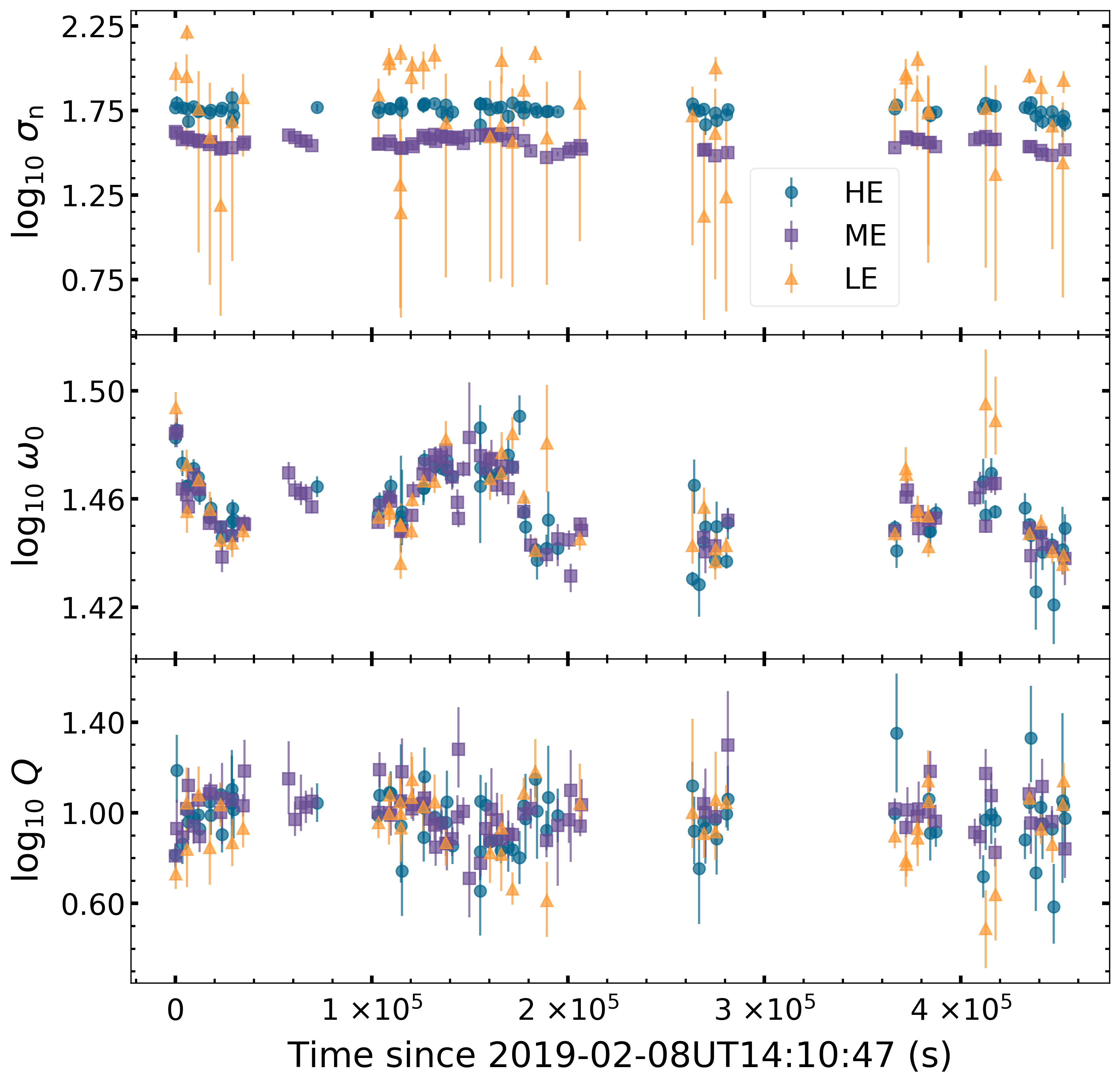}
    \caption{Evolution of part of the parameters in the different energy bands with time when $Q$ is free. From top to bottom: Time evolution of $\sigma_n$, $\omega_0$, and $Q$.}\label{para_time}
\end{figure}

\begin{figure}
    \centering
    \includegraphics[width=0.47\textwidth]{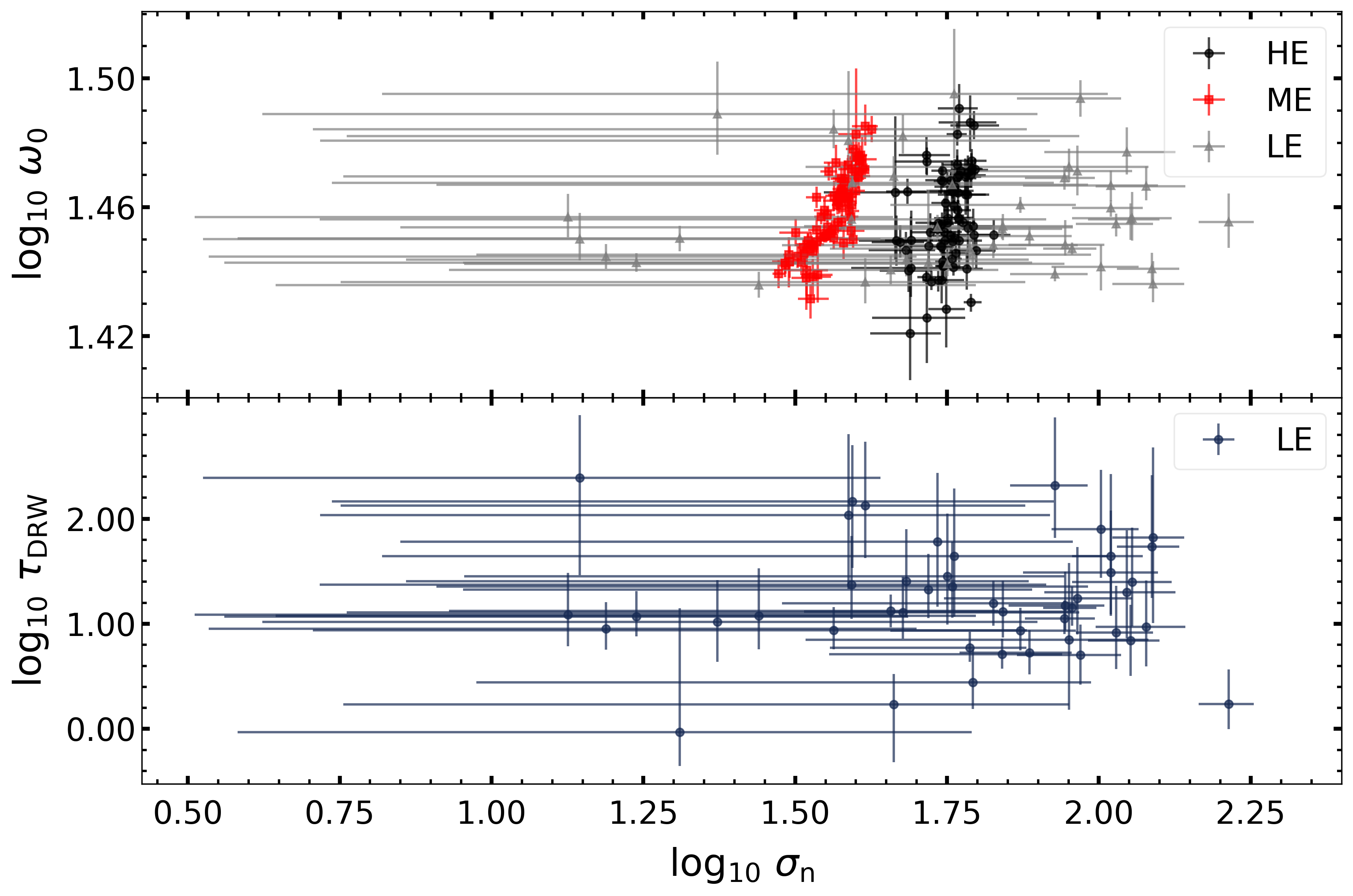}
    \caption{Upper panel: Relationship between parameters $\omega_0$ and $\sigma_n$ in the different energy bands when $Q$ is free. Lower panel: Relationship between parameters $\tau_{\rm DRW}$ and $\sigma_n$ in the LE band when $Q$ is free.}\label{sigma_n}
\end{figure}

To capture enough details from the time series data, 
we chose a time bin of 0.03\,s for all data. A total of 362 good time intervals (GTIs) were fitted, and the states of these time intervals throughout the entire source evolution process were annotated in red on the HID, with each point represents an ExpID\footnote{To reduce the size of a single file, an observation is artificially split into multiple segments, called "exposures", which represent only a time segmentation within an observation, distinct from the traditional exposure concept of a camera or CCD imaging instrument.} (Fig.~\ref{HID}). Most of the GTI durations are approximately in the range of 100\,s to 700\,s. 
In the following section, we show our main results.

\subsection{General GP fitting results}
The DRW and SHO kernels, 
as well as their additive combinations with the AWN term, were tested against the X-ray data.
The additive combination of SHO, DRW, and AWN is found to be the best model to characterize the variability features of the X-ray data.
The AWN term is necessary, as adding it significantly reduces the $\rm AIC_c$ for small sample sizes \citep{1974ITAC...19..716A, burnhamMultimodelInferenceUnderstanding2004}. For example, under the same GTI, the $\rm AIC_c$ values are 108220 for the SHO and DRW, and 108172 for the SHO, DRW, and AWN.

A difference of \( \Delta \mathrm{AIC}_{\mathrm{c}} > 10 \) is generally taken as decisive evidence against the model with the higher \( \mathrm{AIC}_{\mathrm{c}}\) \citep{A_2007}. 
For \( \Delta \mathrm{AIC}_{\mathrm{c}} = 10 \), the Akaike weight \citep[see][]{AKAIKE19813, burnhamMultimodelInferenceUnderstanding2004} is $w=e^{-10/2}\approx0.0067$, meaning that the preferred model is $1/w\approx150$ times better supported.
Interpreted as a likelihood-ratio statistic, \( \Delta \mathrm{AIC}_{\mathrm{c}} = 10 \) roughly corresponds 
to a $3\sigma$ significance level.

The final selected kernel function combines three key components, that is, SHO, DRW, and AWN.
The SHO term serves as the primary tool for period detection, capturing the underlying oscillatory behavior in the data. The DRW component models the stochastic background variability, accounting for correlated noise processes, along with its characteristic timescale. The AWN term handles another white noise component in the data in addition to the measurement uncertainties.

Figs.~\ref{LE_Qfree} -- \ref{HE} show examples of the fitting results for the LE and HE data. The hardness ratio of the data we fit is concentrated around 0.3. The LE light curve has the highest mean count rate (approximately $4500\,\mathrm{cts\ s^{-1}}$), 
followed by HE (approximately $600\,\mathrm{cts\ s^{-1}}$), and ME shows the lowest value (approximately 
$200\,\mathrm{cts\ s^{-1}}$). 

In the fittings, we at first set the quality factor $Q$ in the SHO kernel as a free parameter. This means that we did not have a prior belief about the QPO signature, although QPOs have been reported in the data through the frequency-domain method.
Indeed, the median value of $Q$ is restricted within the range of $5-10$ by the data (Figs.~\ref{LE_Qfree}~and~\ref{HE}),
indicating that the SHO behaves as a good oscillator.
The SHO PSD displays a clear peak at $\omega_0/(2\pi)\approx4.9$\,Hz.
The fitting to the LE light curve yields $\tau_{\rm DRW}\sim5$\,s. 
Looking at the DRW PSD, this characteristic timescale corresponds to the break frequency $f_b$ in the DRW PSD, given by $\tau_{\rm DRW}=1/(2\pi f_b)$.

For each light curve, we find $\sigma_{n} > \sigma_{\rm DRW}$. The LE data have the largest $\sigma_{n}$ and $\sigma_{\rm DRW}$ among the LE, ME, and HE light curves, since the LE data have the highest count rate.
The damping timescale in the DRW kernel is not constrained in the cases of the HE and ME observations (Fig.~\ref{HE}). This means that the stochastic variability is still correlated in the duration of the given GTI.
The AWN power of the HE and ME data dominates over almost all effective frequencies. For the LE data, the DRW power exceeds the AWN power below $\sim f_b$.

We again fit the same data with a fixed $Q=30$ (the $1\sigma$ upper limit obtained above), which ensures a quasiperiodic component in the GP model and reduces the number of free parameters.
The results are shown in Fig.~\ref{LE_Q30}.
The fixed $Q$ results in a significantly smaller $S_0$, which arises from the intrinsic degeneracy between $S_0$ and $Q$ as discussed in Sect. 2. Given a fixed centroid frequency, a larger $Q$ inherently corresponds to a smaller $S_0$. 
The oscillation frequency and the parameters describing the background variability (DRW and AWN components) are nearly unchanged from a free $Q$ to a fixed $Q=30$. This means that the QPO can be modeled as a high-quality oscillation.

In the PSD plot, the blue line in the background represents the Lomb-Scargle periodogram (LSP), which is used as a supplementary method for comparison with the GP PSD. 
The LSP is a generalized form of the discrete Fourier power spectrum that is exactly equivalent to the least-squares fitting of a sinusoid at each trial frequency. 
Unlike the classical fast Fourier transform (FFT), the LSP is believed to remain statistically correct when the observations are uneven in time and when the data have point-by-point errors\footnote{See \cite{osullivanModellingStochasticQuasiperiodic2024} for caveats when applying the LSP to unevenly sampled data.} \citep[][]{2018ApJS..236...16V}. 
Technically, the LSP PSD was obtained by directly applying the \texttt{LombScargle} function from the \texttt{astropy} package in Python.
The GP approach offers a distinct advantage by decomposing the PSD into three physically interpretable components.
This allows us to distinguish among the different physical processes contributing to the observed variability.

Figure~\ref{good_pow} shows the representative PSDs\footnote{The data used are from the LE band with the ExpID P021400201101.} calculated by GP and obtained using the \texttt{powspec} tool.
Combining the GP results, we can now dive into the \texttt{powspec} PSD (frequency-domain method).
The SHO component corresponds to the QPO signal (green peak in the right panel of Fig.~\ref{good_pow}). 
The DRW term matches the red line in the right panel of Fig.~\ref{good_pow}, both of which describe the PSD index variation in the low-frequency range. 
As mentioned above, the DRW component of the LE 
light curve is different from that of ME and HE light curves with a larger $\sigma_{\rm DRW}$ and a smaller $\tau_{\rm DRW}$, causing the DRW component to exceed the AWN component below $\sim1/(2\pi \, \tau_{\rm DRW})$\,Hz. This also explains why the LE \texttt{powspec} PSD deviates from white noise below $1\,$Hz.
We notice that $\tau_{\rm DRW}$ is constrained within the range of $[10\times t_{\rm binsize},0.1\times t_{\rm baseline}]$ by the LE data, where $t_{\rm baseline}$ is the duration of the given GTI.
This indicates that $\tau_{\rm DRW}$ is well constrained.
The frequency-domain method reveals the presence of a second harmonic component at $\sim 9$\,Hz \citep{2020MNRAS.499..851Z}, which cannot be captured by the GP method.

In GP method, the SHO term characterizes the QPO signal, and the rms of the SHO component is given by Eqs.~(\ref{varsho})~and~(\ref{SHO_newpara}), i.e., $\sqrt{S_0 \omega_0 Q}$.
The fractional rms of the SHO component is defined as
\begin{equation}
    \text{fractional rms} = \frac{\sqrt{S_0 \omega_0 Q}}{\text{mean count rate}}.
\end{equation}

The fractional rms of the SHO component for each GTI is derived from the GP fitting results, as shown in Fig.~\ref{rms}. 
The colors (purple, yellow, and red) correspond to the three discrete energy bands, with each line indicating the rms value for that GTI within the respective band.
Specifically, the fractional rms obtained via the GP method ranges from $1.7-3.3$\% in the LE band (1--10\,keV), increases to $7.5-13.6$\% in the ME band (10--30\,keV), and then decreases to $4.2-9.9$\% in the HE band (30--100\,keV). For comparison, the black lines in Fig.~\ref{rms}, with values of $2.7$\%, $13.3$\%, and $7.2$\% for the LE, ME, and HE bands, respectively, represent the fractional rms values calculated using the frequency-domain method for a representative ExpID\footnote{The ExpID used is P021400201101.}. These values lie within the corresponding fractional rms ranges inferred from the GP fits, indicating good consistency between the two approaches \citep{liuTransitionsOriginTypeB2022}.

\subsection{Statistical analysis of the GP results}

Based on the GP fitting results for the 362 GTIs, 
we performed a statistical analysis of the parameter values for the significant type-B QPO observations (204 GTIs). 
Figure~\ref{para_time} shows the evolution of the kernel parameters as a function of time and energy. 
The amplitude of the AWN term $\sigma_n$ has larger uncertainties in some LE GTIs.
This is because, on long time scales, the DRW component has a larger variance than the AWN term, which is clearly visible in the PSD plot.
The values of $\sigma_n$ obtained from the LE data are generally larger than those from the ME and HE bands. Meanwhile, the $\sigma_n$ values in the ME and HE bands are also well constrained and show little temporal variation.

The natural oscillation frequency $\omega_0$ in the SHO kernel varies from $26\rm\,rad/s$ (4.1\,Hz) to $31\rm\ rad/s$ (5.0\,Hz) over time. The quality factor $Q$ varies from $4$ to $20$.
In general, $\omega_0$ and $Q$ obtained from the ME data have the smallest uncertainties.
Note that the ME light curve has the lowest count rate among the three energy bands and also the smallest well-constrained $\sigma_n$.

The relation between $\sigma_n$ and $\omega_0$ is plotted in the upper panel of Fig.~\ref{sigma_n}. There is a positive trend in the ME cases.
However, the dynamic ranges of both variables are small. Consequently, we lack sufficient statistical confidence to confirm the validity of this trend. 
The lower panel of Fig.~\ref{sigma_n} illustrates the dependence of $\sigma _n$ and $\tau_{\rm DRW}$ of the LE band, since only the LE data have the well-constrained $\tau_{\rm DRW}$. The mean value of LE $\tau_{\rm DRW}$ is $16$\,s, with a standard deviation of $1$\,s. The $\sigma _n$ and the $\tau_{\rm DRW}$ show no apparent correlation.

\subsection{Key findings}

Through GP modeling, we successfully 
decomposed the light curve of this source and identified its variability components. 
We also systematically compared their evolution across both energy and temporal domains. 
Our key findings are as follows.

\begin{enumerate}
    
    \item[-] The stochastic variation component (DRW) of the LE band is markedly different from that of the ME and HE bands. In LE, correlations on the scale of $\tau_{\rm{DRW}}$ were lost, whereas the data for ME and HE remain correlated on the timescale of the length of the light curve.
   
    \item[-] The SHO component behaves as a high-quality oscillator and does not exhibit obvious evolution in either the energy domain or the time domain, but the SHO component in the ME has the smallest uncertainty, which may be related to its smallest $\sigma_n$. Also, the ME data have the largest fractional rms, while the HE data show moderate values and the LE data the lowest. 

    \item[-] An AWN term is needed, which implies that there is an extra uncorrelated (white noise) variation in the data.
        
\end{enumerate}

These results contribute to a better understanding of the origins of the different components and their interactions.

%%%%%%%%%%%%%%%%%%%%%%%%%%%%%%%%%%%%%%%%%%%%%%%%%%%%%%%%%%%%%%%%%%%%%%%%%%

\section{Discussion}
Using the GP method, we analyzed Insight-HXMT observations of MAXI J1348--630 during its type-B QPO phase. The GP method is a time-domain approach that models the light curve directly by superposing kernels, each with a specific mathematical form and physical interpretation. 
This is different from the frequency-domain method which typically fits the PSD with multiple Lorentzian components.\par

For MAXI J1348--630, during its type-B QPO phase, three distinct variability components are identified using the GP method: one quasiperiodic component (the SHO term) and two stochastic components (the DRW term and the AWN term). The SHO term corresponds to the Lorentzian feature that represents the QPO signal in the frequency domain. 
The DRW term captures the low-frequency red noise seen in the PSD, and the AWN term structurally represents an extra white noise component in the data. In frequency-domain analyses, such an extra white noise component has not received particular attention. However, in our case, we find that it dominates the data and likely carries important physical significance.\par

By leveraging the physical interpretability of GP kernels, we can explore the physical origins of the corresponding variability components. In the posterior distributions of the model parameters, the characteristic timescale $\tau_{\rm DRW}$ shows distinct and intriguing behavior across the energy bands. Specifically, $\tau_{\rm DRW}$ is well constrained in the LE band with typical values around $\sim 10$\,s, but is poorly constrained in both the ME and HE bands. 
The average lengths of the ME and HE GTI light curves are $\sim 400$\,s and $\sim 300$\,s, respectively.
This lack of constraint on $\tau_{\rm DRW}$ in ME and HE suggest that the physical relaxation timescales at higher energies are intrinsically longer than those at lower energies.\par

The distinct variation of $\tau_{\rm DRW}$ across different energy bands are likely connected to the different physical origins of the X-ray photons. The LE photons primarily originate from the accretion disk, while the ME and HE photons are mainly produced through the inverse Compton scattering of disk photons in the corona. 
The DRW component could be associated with thermal instability in the accretion disk and the corona.
The relaxation timescale was established by the balance of heating and cooling in the fluid.
The shorter $\tau_{\rm DRW}$ in the LE data implies that the thermal instability timescale in the accretion disk is shorter than that in the corona.
The thermal instability timescale is given by $t_{\rm th}\sim (H/R)^2\, t_{\rm vis}$, where $t_{\rm vis}$ is the viscous timescale for the radial redistribution of mass and angular momentum, and $H/R$ is the
ratio between the scale height $H$ and the radial extent $R$ of the matter structure \citep{shakuraTheoryInstabilityDisk1976}. If we set $t_{\rm th}\sim \tau_{\rm DRW}\sim10\,$s, taking $H/R\sim0.05$, we obtain $t_{\rm vis}\sim4000\,$s.
This is reasonable for a stellar black hole system.
In contrast, $\tau_{\rm DRW}$ in the ME and HE bands cannot be well constrained by the current GTI light curves. This implies that the characteristic timescales in these bands are at least on the order of, or longer than, the duration of the GTIs, i.e., several hundred seconds or more. This is consistent with the expectation that the jet-like corona can extend up to hundreds of $R_{\rm g}$ during the type-B QPO phase \citep[][]{garciaTwocomponentComptonizationModel2021}.\par

Furthermore, the characteristic frequency of the DRW PSD is defined as $f_b = 1/(2\pi\,\tau_{\rm DRW})$, indicating that a longer $\tau_{\rm DRW}$ in the ME and HE bands corresponds to a lower characteristic frequency in these energy bands. This energy-dependent trend of $f_b$ is consistent with Insight-HXMT observations of MAXI~J1820+070, where $f_b$ remains approximately constant at 0.1\,Hz below 30\,keV but decreases to below 0.04\,Hz above 30\,keV. \citet{Gao:2024bol} attribute this behavior to the origin and scattering number of seed photons. Specifically, photons originating from the opposite side of the accretion disk traverse through the corona, undergoing multiple Compton upscatterings before reaching the observer. The number of scatterings is determined by the photons' paths through the corona, and the emission direction tends to align with the photon's incident direction into the corona when the optical depth exceeds 4 \citep{pozdnyakovComptonizationShapingXray1983}. Lower-frequency photons originate from more distant disk regions, where photons undergo more scatterings in the corona, resulting in higher-energy emission and a lower characteristic frequency.

For MAXI~J1348--630, spectral-timing analysis suggests the presence of two distinct coronae \citep{garciaTwocomponentComptonizationModel2021, zhangPeculiarDiskBehaviors2022}. The smaller corona is nearly spherical, with a characteristic size on the order of tens of $R_{\rm g}$, while the larger corona exhibits a jet-like shape extending to scales of several hundred $R_{\rm g}$, as inferred from the time-dependent {\tt vKompth} model \citep{garciaTwocomponentComptonizationModel2021, alabartaGeometryComptonizationRegion2025a}. Based on this geometry, we propose that seed photons with higher characteristic frequencies ($f_b$) in the lower energy bands originate from regions closer to the black hole and are predominantly Comptonized in the spherical small corona, where the photon path lengths are relatively short. In contrast, seed photons with lower $f_b$ observed at higher energies likely arise from more distant regions of the disk and undergo Comptonization in the extended, jet-like corona, resulting in longer photon path lengths, particularly considering the low inclination (${29.3^{+2.7}_{-3.2}}^{\circ}$) of MAXI~J1348--630 \citep{carotenutoModellingKinematicsDecelerating2022}. Moreover, the optical depth remains below 2 during the type-B QPO phase \citep{garciaTwocomponentComptonizationModel2021}. This may indicate that the seed photons originate not only from the far side of the disk, as proposed for MAXI~J1820+070, but potentially from the entire disk region.\par

Interestingly, an AWN component is needed to fit the LE, ME, and HE light curves. We could interpret this AWN component as a second DRW process with a characteristic timescale so short that it remains unresolved by our 0.03\,s binning \citep[e.g.,][]{Zhang_2025}. We also attempted to use the light curve in a time bin of 0.01\,s and replace the AWN term with a second DRW component to fit the data, but the short damping timescale is still not detected. This indicates that the second DRW component has a characteristic timescale of $\tau_{\rm DRW} < 0.01\,$s.\par

\citet{Gao:2024bol} reported 
the broadband PSD of the multiband X-ray data for MAXI~J1820+070.
Looking at their Fig.~2, we find that 
the broadband PSD (not including the QPO peak) can be described with a broken power-law:
\begin{equation}
P(f) \propto f^{-\alpha_i} \quad
\left\{
\begin{array}{ll}
\alpha_1, & f < f_{\mathrm{b},1} \\
\alpha_2, & f_{\mathrm{b},1} < f < f_{\mathrm{b},2} \\
\alpha_3, & f > f_{\mathrm{b},2},
\end{array}
\right.
\end{equation}
where $\alpha_i$ is the PSD index and $f_{\mathrm{b},i}$ is the break frequency. 
For instance, the 1--10\,keV PSD index of MAXI~J1820+070 typically varies from $\alpha_1\approx0$ to $\alpha_2\approx1$ and then to $\alpha_3\approx2$, with the two break frequencies being $f_{\mathrm{b},1}\sim 0.06$\,Hz and $f_{\mathrm{b},2}\sim 2$\,Hz. The two break frequencies reflect two timescales, given by 1/$(2\pi f_b)$, of 3\,s and 0.08\,s, respectively.
The shorter timescale decreases to $\sim 0.02$\,s for the data above 76\,keV. Such a broadband PSD has also been reported in other sources with $f_{\mathrm{b},2}\sim5$\,Hz, for example, XTE~J1859+226 \citep[][]{mottaBlackHoleMass2022} and GX~339--4 \citep[][]{zhangSystematicStudyHighfrequency2023}.
For MAXI~J1348--630, we identify two characteristic timescales: $\sim10$\,s and $<0.01$\,s. The different physical timescales reflect distinct physical environments in these sources.\par

Physically, the ANW, that is, the rapid DRW variability with a damping timescale less than 0.01 s appearing in all the three energy bands, likely points to small-scale, high-frequency instabilities, most plausibly magnetorotational instability (MRI) operating in both the accretion disk and the corona. This suggests that their magnetic fields are dynamically coupled. Moreover, since $\sigma_{n}>\sigma_{\rm DRW}$, the MRI-driven fluctuations exhibit a larger amplitude than the slower thermal instabilities.\par
 
We also detect a quasiperiodic signal (the SHO term) at similar frequencies ($\sim 4.6$\,Hz) and comparable quality factors across all three energy bands. 
This common periodicity implies a shared driving mechanism and a similar emission-region structure across the disk and the corona. This also suggests that the disk and corona are coupled on a longer timescale of $2\pi/\omega_0\approx0.2 $\,s.
This disk-corona oscillation could be driven by a relative large-scale modulation of the MRI.\par

Our analysis reveals a timescale hierarchy as follows:
the AWN (rapid DRW) has the shortest timescale; followed by the SHO; then the disk DRW; and finally the corona DRW. 
The two shorter timescale terms, the AWN and the SHO, may arise from MRI turbulence and therefore appear in all three energy bands (LE, ME, and HE), 
linking the disk and corona through shared magnetic loops. 
The two longer DRW timescales possibly trace thermal-viscous relaxation: the smaller, well-resolved timescale in LE matches the thermal instability timescale of the thin, gas-pressure-dominated disk, whereas the currently unconstrained damping timescale in ME and HE points to a longer thermal instability timescale in the geometrically thick, Comptonizing corona.
However, as shown in Fig.~\ref{sigma_n}, the three processes (SHO, DRW, and AWN) appear to be independent and have no influence upon each other.\par

At present, the \texttt{celerite} framework is confined to the kernels that can be written as finite sums of exponentials. This restriction makes it difficult to capture subtle features in light curves, such as the variability whose PSD follows an $f^{-1}$ index within a specific frequency band \citep[e.g.,][]{Gao:2024bol}.\par

%%%%%%%%%%%%%%%%%%%%%%%%%%%%%%%%%%%%%%%%%%%%%%%%%%%%%%%%%%%%%%
\begin{acknowledgements}
      The authors sincerely thank the referee for the helpful suggestions that have significantly improved the quality of our work.
      This work used data from the Insight-HXMT mission, a project, funded by the China National Space Administration and the Chinese Academy of Sciences.
      RM acknowledges support from the Royal Society Newton Funds.
      DY thanks the funding support from the National Natural Science Foundation of China (NSFC) under grant No. 12393852.
\end{acknowledgements}

\bibliographystyle{bibtex/aa}
\bibliography{bibtex/ref}

\end{document}